\documentclass[conference]{IEEEtran}
\IEEEoverridecommandlockouts
\usepackage{cite}
\usepackage{amsmath,amssymb,amsfonts}
\usepackage{accents}
\usepackage[linesnumbered,ruled]{algorithm2e}
\usepackage[noend]{algpseudocode}
\usepackage{graphicx}
\usepackage{textcomp}
\usepackage{booktabs}
\usepackage[table,xcdraw]{xcolor}
\usepackage{xcolor}
\usepackage{comment}
\usepackage{multicol,multirow}
\usepackage[T1]{fontenc}
\def\BibTeX{{\rm B\kern-.05em{\sc i\kern-.025em b}\kern-.08em
    T\kern-.1667em\lower.7ex\hbox{E}\kern-.125emX}}
\usepackage{todonotes}
\usepackage{caption}
\newsavebox{\measurebox}
\usepackage{threeparttable}
\usepackage[font=small]{caption}
\usepackage{enumitem}
\usepackage{subfigure}
\usepackage{dblfloatfix}

\usepackage{hyperref}
\hypersetup{
    colorlinks=false
}

\newcommand{\cheng}[1]{{{\color{blue}#1}}{}}

\newcommand*{\vsepfbox}[1]{%
  \begingroup
    \sbox0{\fbox{#1}}%
    \setlength{\fboxrule}{0pt}%
    \mbox{\kern-\fboxsep\fbox{\unhbox0}\kern-\fboxsep}%
  \endgroup
}

\begin{document}

\title{Safety Interventions against Adversarial Patches in an Open-Source Driver Assistance System}

\author{
\IEEEauthorblockN{{\fontsize{10.5}{12.6}\selectfont 
Cheng Chen, Grant Xiao\IEEEauthorrefmark{2}, Daehyun Lee \IEEEauthorrefmark{3}, Lishan Yang\IEEEauthorrefmark{3}, Evgenia Smirni\IEEEauthorrefmark{4}, Homa Alemzadeh \IEEEauthorrefmark{2}, Xugui Zhou}}
\IEEEauthorblockA{
    Louisiana State University, Baton Rouge, LA 70803 \{cchen72, xuguizhou\}@lsu.edu \\
    \IEEEauthorrefmark{2}{University of Virginia},
    Charlottesville, VA 22904 
    \{yeq8mf, ha4d\}@virginia.edu \\
    \IEEEauthorrefmark{3}{George Mason University},
    Fairfax, VA 22030
    \{dlee217, lyang28\}@gmu.edu \\
    \IEEEauthorrefmark{4}{William \& Mary},
    Williamsburg, VA 23187
    \{esmirni\}@cs.wm.edu
    }
}

\maketitle

\begin{abstract}
Drivers are becoming increasingly reliant on advanced driver assistance systems (ADAS) as autonomous driving technology becomes more popular and developed with advanced safety features to enhance road safety. However, the increasing complexity of the ADAS makes autonomous vehicles (AVs) more exposed to attacks and accidental faults. In this paper, we evaluate the resilience of a widely used ADAS against safety-critical attacks that target perception inputs. Various safety mechanisms are simulated to assess their impact on mitigating attacks and enhancing ADAS resilience. Experimental results highlight the importance of timely intervention by human drivers and automated safety mechanisms in preventing accidents in both driving and lateral directions and the need to resolve conflicts among safety interventions to enhance system resilience and reliability. [Code Available at {https://doi.org/10.6084/m9.figshare.28691090}]

\end{abstract}


\section{Introduction}

Advanced driver assistance systems (ADAS) offer $\text{Level-2}$ autonomous driving features \cite{SAEroadmap} such as automatic lane centering (ALC) and adaptive cruise control (ACC). They are equipped on over 300 million passenger vehicles worldwide~\cite{adasmarket} and are estimated to reach a market value of USD 73.74 billion by 2031 \cite{adasmarketvaule}. 
This proliferation stems from the rapid advances in sensing and computing technologies and machine learning, which also raises concerns about the safety of vehicles and human drivers due to the increasing complexity and connectivity of ADAS.
For example, recent studies show that Tesla's vehicles have the highest fatal accident and crash rates among all car brands in the U.S.~\cite{taslaAccidents2,usdot_lv2}. 

The core features of ADAS are ACC and ALC, which autonomously regulate a vehicle’s speed and steering angle to maintain lane centering and a safe following distance from lead vehicles. To achieve these functionalities, ADAS relies on deep learning (DL) models to detect leading objects and lane markings from camera inputs. However, DL models are inherently vulnerable to input perturbations \cite{jha2019ml,jha2020ml, zhou2022robustness, zhou2024runtime}, and failures in object or lane line detection can lead to catastrophic safety risks \cite{sato2021dirty,zhou2024runtime}. Consequently, studying the resilience of ADAS to camera input perturbations
is critically important, particularly for vision-centric systems like Tesla Autopilot \cite{autopilot} and Subaru EyeSight \cite{eyesight}.

Previous work has shown DL-based ADAS are vulnerable to perception attacks, including digital attacks that manipulate live camera feeds by directly compromising control software \cite{jia2019fooling} and physical attacks employing adversarial stickers on road signs \cite{eykholt2018robust}, the road surface \cite{tencent2019experimental}, or camera lenses \cite{li2019adversarial}, as well as adversarial patches projected or printed on vehicles ahead \cite{ma_wip_2023,sato2021dirty}.
However, these works largely overlook the role of human-driver intervention and the impact of integrated safety mechanisms within the ADAS control loop, such as advanced emergency braking systems (AEBS) and lane departure warnings.
These advanced safety features, designed to enhance road safety, are increasingly standard in modern ADAS and are critical to evaluating the resilience of ADAS.

While some studies have evaluated safety features like automatic emergency braking (AEB), these efforts are often limited to component-level analysis \cite{ma2021sequential} without systematically testing the safety mechanisms and ADAS as a whole. Others focus narrowly on a single feature, such as ACC \cite{zhou2024runtime}, neglecting the interdependencies among ADAS functionalities. Recent advancements in machine learning have shown promise in enhancing the resilience of robotic vehicles against sensor attacks \cite{Choi2020software, dash2021pid, dash2024specguard}. However, these methods have yet to be evaluated on commercial ADAS. 

To fill this gap, we systematically assess the resilience of a widely used open-source commercial ADAS, OpenPilot~\cite{openpilot}, against perception attacks (adversarial patches) while considering the interventions of existing safety mechanisms (such as AEB) and human drivers. 
Given that real-world road tests require significant time and resources and pose substantial risks to human safety and vehicle integrity during collision scenarios, especially when studying safety-critical attacks that could lead to crashes, we have developed a simulation platform for realistic experimental studies. Our platform integrates real-world control software, a physical-world simulator, a driver reaction simulator, and key ADAS safety mechanisms (AEB and FCW), all implemented in accordance with international standards and government-issued transportation guidelines.

The main contributions of the paper are as follows:
\begin{itemize}
    \item Analyzing the effectiveness of basic ADAS safety interventions in mitigating the adversarial effect of physical patches on the ALC and ACC operation.
\item Developing an open-source platform for realistic simulation of attacks and different levels of safety intervention. 
\item Studying the conflicts or coordination among different safety interventions, including automated mechanisms and human drivers.
\item Comparing basic safety mechanisms with an automated ML-based mitigation method.

\end{itemize}

Our study demonstrates that OpenPilot is highly vulnerable to adversarial patch attacks targeting ACC and ALC, often failing to detect the front vehicle at close range and exhibiting unsafe, aggressive speed control even in benign conditions, such as applying a hard brake only at a very close distance when approaching the lead vehicle. The findings highlight that basic safety mechanisms and human intervention may be more effective than certain ML-based automated mitigation methods in preventing accidents across both driving and lateral directions in the simulated scenarios. Lateral attacks, in particular, remain challenging to mitigate, though highly alert drivers achieve notably better prevention rates. Furthermore, the results underscore the critical role of independent sensors for AEBS, the potential of AEB to prevent lateral accidents, and the pressing need to resolve conflicts among safety features to enhance overall system reliability and resilience.

\section{Preliminaries}
\subsection{Advanced Driver Assistance System}

Advanced Driver Assistance Systems (ADAS) are technologies now widely used in vehicles to enhance road safety and driving comfort by freeing up driver attention and reducing human error. With a number of assistive features and semi-autonomous driving designs, it effectively enhances driving safety and driver comfort over long distances. ADAS systems rely on sensors and cameras to collect real-time data about the vehicle's surroundings, which can be used to enable predictive action and alert the driver to potential hazards.

\begin{figure}[t]
    \centering
    \includegraphics[width=\linewidth]{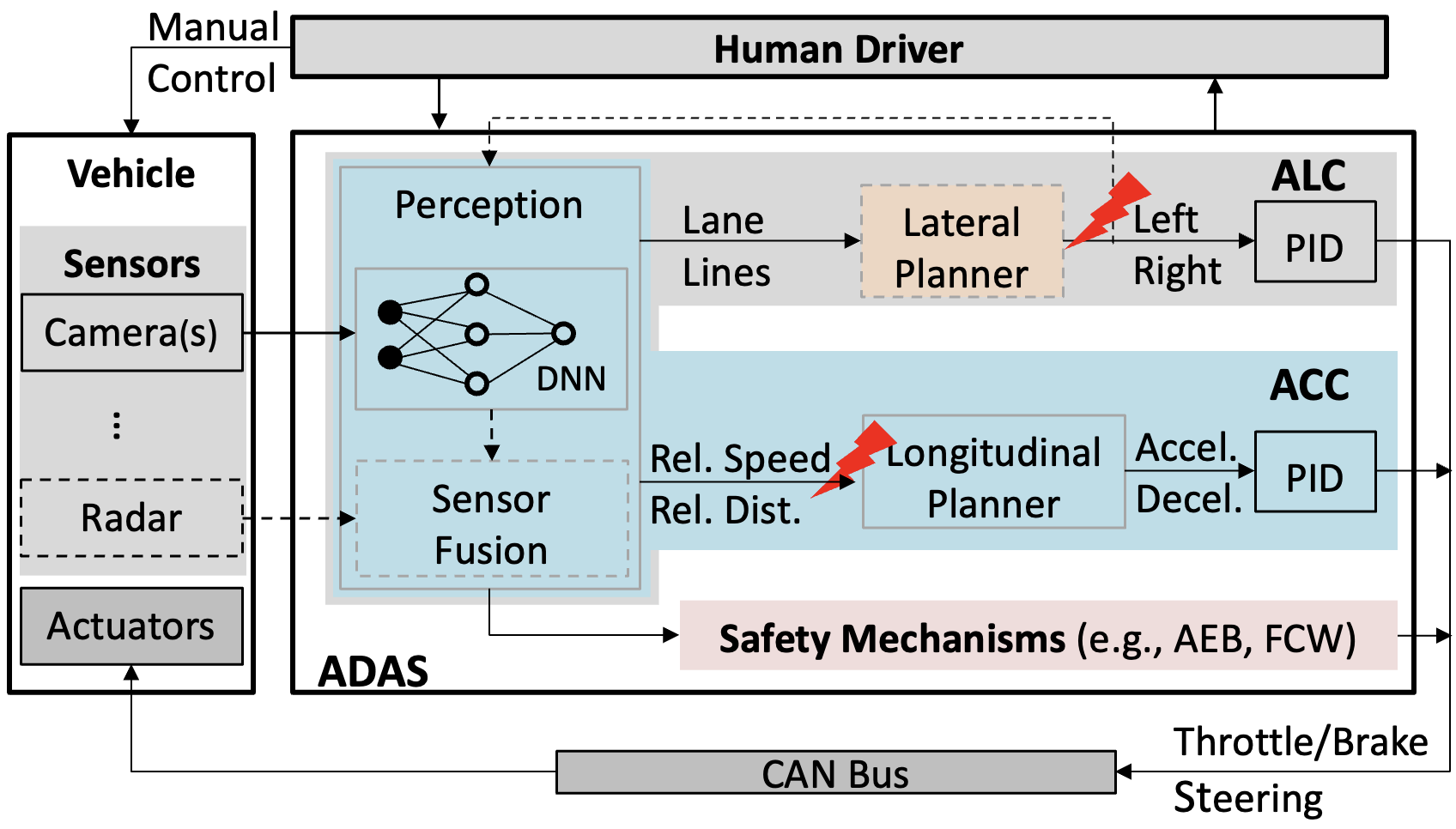}
    \caption{Overview of the Control Structure of an ADAS.}
    \label{fig:adasoverview}
    \vspace{-1em}
\end{figure}

Fig. \ref{fig:adasoverview} shows the control structure of a typical ADAS. Key ADAS functions include ACC, ALC, AEB, Blind Spot Detection, Traffic Sign Recognition, and Driver Monitoring Systems, among others. ADAS technologies are designed to automate driving tasks by gradually automating driving tasks and ultimately achieving fully automated driving functions. These systems significantly improve safety by reducing the number of accidents caused by human error, while increasing the overall comfort and convenience of driving. As technology advances, ADAS integration in vehicles is becoming more common, resulting in smarter and safer roads.
    
\textbf{Autonomous Driving Levels: }Autonomous driving technologies are categorized into different levels according to the vehicle's automation capabilities and the need for human intervention, ranging from Level 0 (no automation) to Level 5 (full automation). Current commercial ADAS operate at Level 2, where human drivers must continuously monitor the driving environment and ensure safety at all times \cite{SAEroadmap}.

\textbf{Safety Mechanisms: }Safety mechanisms are an integral part of ADAS to prevent accidents and protect vehicle occupants through a range of functions.

AEB is an important safety feature that automatically activates the brakes when it detects an impending collision, sometimes before the driver can react. This system is particularly effective at low speeds and in urban driving environments and can significantly reduce rear-end accidents.
Closely working with AEB, Forward Collision Warning~(FCW) monitors the road ahead with a forward-facing camera and radar, and when it detects a risk of a collision ahead, the system warns the driver to take evasive action. In many cases, the FCW system can warn the driver several seconds in advance, giving enough time to adjust speed or avoid obstacles.

These mechanisms mark a milestone in automotive safety, improving driving and road safety and paving the way for further ADAS innovations.

    
\textbf{Openpilot: }OpenPilot \cite{openpilot} is an open-source ADAS driving software from Comma.ai, Inc. When the OpenPilot software runs on Comma.ai's hardware and connects to a car, it becomes an autonomous driving system that can control the car, allowing for upgrades to original systems that lack ADAS features. 
OpenPilot supports more than 300 vehicle models \cite{Carmodels-OpenPilot} and has over 10,000 users, with more than 100 million miles driven. OpenPilot utilizes a system-level end-to-end design that uses DL models to predict the information necessary to avoid risks and plan the car's trajectory, based directly on images captured by the camera on the Comma hardware. This is a departure from traditional self-driving solutions, which rely on separate units of perception, prediction, and planning that operate in conjunction with each other.

\subsection{MetaDrive Driving Simulator}

MetaDrive \cite{li2021metadrive} is an open-source simulator platform for autonomous driving research, specifically designed to provide an efficient and scalable simulation environment. It is the official simulator for OpenPilot by Comma.ai. MetaDrive provides a controlled-risk environment for researchers to test and optimize autonomous driving algorithms, including perception, decision making, and motion control strategies. 
We do not use CARLA \cite{Dosovitskiy17}, another widely used physical-world simulator, due to its incompatible Python environment with the OpenPilot (v0.9.7) control software.

\section{Approach}
\subsection{Threat Model}
\label{sec:attackmodel}
We assume the attacker has the capability to launch efficient physical-world attacks on the camera inputs of ADAS by introducing adversarial patches on the rear of a lead vehicle (LV)~\cite{hoory2020dynamic} or directly on the road \cite{sato2021dirty} with the goal of compromising the ACC and ALC functionalities. These capabilities may include prior information about the victim vehicle’s driving path and full knowledge of the target ADAS (e.g., ALC and ACC) implementation details through reverse-engineering a rented or purchased vehicle with the same model~\cite{tencent2019experimental} or researching open-source ADAS materials \cite{openpilot}. 
The assumption of attack capability is realistic and similar to previous work in the literature \cite{sato2021dirty,hoory2020dynamic,chahe2023dynamic,man2023person,lou2024first,zhou2024runtime}.
The resulting perturbations compromise the ADAS perception module (see Fig. \ref{fig:adasoverview}), leading to incorrect predictions of critical information, such as the lead vehicle’s distance from the ego vehicle (EV) and lane line positions. Consequently, these errors propagate to the ADAS control software, causing malfunctions in the ALC and ACC modules, which can result in safety hazards or accidents, such as collisions or lane departures.

In this paper, we focus on analyzing the impact and effectiveness of existing safety mechanisms in mitigating physical-world perception attacks.

\subsection{Attack Design and Implementation}

In this paper, we examine attacks targeting ALC and ACC features of ADAS.

\begin{figure}[t]
    \centering
    \includegraphics[width=.65\linewidth]{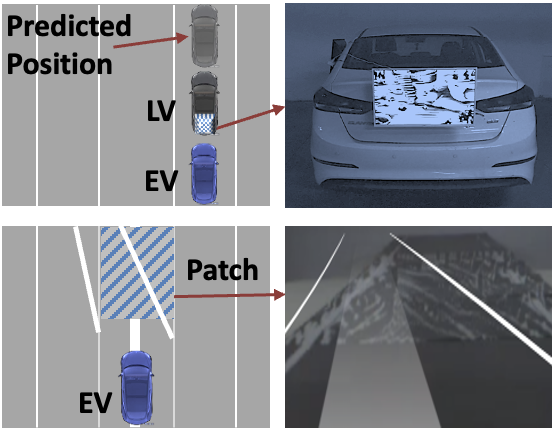}
    \caption{Example physical attacks against ACC and ALC by adding an adversarial patch on the rear of the lead vehicle (LV) \cite{hoory2020dynamic} or on the road \cite{sato2021dirty}.}
    \label{fig:faulttype}
    \vspace{-1em}
\end{figure}

For ALC attacks, the attacker aims to deviate the vehicle’s driving direction by altering predictions of lane lines or desired curvature. Prior work has demonstrated the feasibility of physical adversarial perturbations on camera inputs to achieve this goal \cite{sato2021dirty}. Here, we adopt a method that compromises ALC functionality by deploying a well-crafted patch on the ego vehicle’s driving path \cite{sato2021dirty} (see Fig. \ref{fig:faulttype}). The attack is activated when the ego vehicle drives over the area containing the patch.

For ACC attacks, we utilize a method explored in previous studies \cite{guo2023adversarial,zhou2024runtime,hoory2020dynamic}, which involves displaying or projecting a physical patch on the rear of a lead vehicle (see top images in Fig. \ref{fig:faulttype}). This attack is triggered when the patch is detected by the ego vehicle within an effective range (e.g., less than 80 meters). It disrupts the prediction of the lead vehicle’s distance, a critical input for the ACC decision-making module.

Note that only a portion of the adversarial patches lead to DNN (deep neural network) misprediction and further propagate to cause unsafe driving behaviors. To better evaluate the impact of safety interventions against these adversarial patches, we directly emulate the effect of the patches by injecting attacks into the DNN output and getting the range of attack values of mispredictions corresponding to adversarial patches from previous work \cite{sato2021dirty,zhou2024runtime}.

\subsection{Safety Mechanisms Design}
\label{sec:safety}
Passenger vehicles are increasingly equipped with advanced safety mechanisms such as AEB, FCW, lane departure warning, and firmware safety checking. In this paper, we implement three-level safety mechanisms to realistically assess the impact of these existing safety mechanisms on improving ADAS resilience to perception attacks (adversarial patches) and driving safety, including basic-level AEBS, application-level safety checking, and human-level driver interventions.

\textbf{AEBS: }
We adhere to the established guidelines and regulations for AEBS \cite{schram_implementation_nodate,http://data.europa.eu/eli/reg/2020/1597/oj,noauthor_grva-12-50r1epdf_nodate} and implement a time-to-collision-based phase-controlled AEBS, following the approach outlined in previous studies \cite{alsuwian_autonomous_2022, zhou2024runtime}. 
Specifically, the AEBS operates by calculating the time to collision (TTC) between the ego vehicle and the lead vehicle, defined as the ratio of Relative Distance (RD) to Relative Speed (RS):
\begin{equation}
ttc = {RD}/{RS} \label{eq:ttc}
\end{equation} 
To determine the braking response, the system first estimates the time required for the ego vehicle to stop under human driver braking:
\begin{equation}
    T_{stop}  = V_{Ego}/a_{driver} \label{eq:tstop}
\end{equation}
where, $V_{Ego}$ is the ego vehicle’s speed and $a_{driver}$ is the driver’s assumed deceleration.

An FCW is triggered to alert the driver of potential collision risks when the estimated time to collision is insufficient for the driver to respond ($T_{react}$) and stop the vehicle safely.
\begin{equation}
    t_{fcw}  = T_{react} + T_{stop} \label{eq:tfcw}
\end{equation}


In this study, we assume an average human reaction time of 2.5 seconds, as reported in the literature \cite{zhou2022strategic, sato2021dirty,zhou2024runtime}. If the driver fails to react to the FCW alert in time, the AEBS activates phased braking based on speed-dependent TTC thresholds:
\begin{equation}
    t_{pb1}  = \frac{V_{Ego}}{3.8}, \quad
    t_{pb2} = \frac{V_{Ego}}{5.8}, \quad
    t_{fb} = \frac{V_{Ego}}{9.8} \label{eq:tpb}
\end{equation}
Specifically, when $ttc$ falls below the cascade thresholds for first-stage braking time ($t_{pb1}$), second-stage braking time ($t_{pb2}$) and full-force braking time ($t_{fb}$),  the system applies 90\% braking, 95\% braking and full braking, respectively. 
The activation sequence of FCW alerts and corresponding AEBS actions are shown in Table \ref{table:AEBS Action}.

\begin{table}[h]
\caption{AEBS Actions.}
\label{table:AEBS Action}
\centering
\begin{tabular}{lllll}
\toprule
    \textbf{TTC} & [$t_{fcw}$,$t_{pb1}$] & [$t_{pb1}$,$t_{pb2}$] & [$t_{pb2}$,$t_{fb}$] & [$t_{fb}$,0] \\ \midrule
    \textbf{Action} & FCW Alert & 90\% Brake & 95\% Brake & 100\% Brake \\ \bottomrule
\end{tabular}
\end{table}

Note that in some car models, the AEBS may also rely on compromised DNN predictions as inputs, potentially degrading its functionality, whereas other models might be equipped with separate sensors to ensure reliable AEBS operation \cite{zhou2024runtime}. Furthermore, certain vehicles may completely disable AEBS functionality when ADAS is activated \cite{Carmodels-OpenPilot}.
To account for these variations and realistically simulate AEBS implementation on actual passenger cars, we design the AEBS mechanism to operate under three distinct configurations: (1) AEBS is disabled, (2) AEBS is activated but relies on compromised sensor data, and (3) AEBS is activated and utilizes inputs from an independent, secure data source.

\textbf{Firmware Safety Checks:}
In addition to safety checking in the control software, OpenPilot also implements firmware safety checking on the output control commands through a CAN interface device, PANDA \cite{Panda}, which is a universal OBD adapter developed by Comma.ai, providing access to almost all vehicle sensors over the CAN bus. However, this PANDA safety checking is not available in the simulation.
To align our safety mechanism with real-world implementations, we replicate the logic from PANDA and design a software-based safety constraint checker that detects if command values are within a predefined safe range, thereby blocking unsafe control commands. For example, to prevent potential hazards, the maximum acceleration and deceleration of the vehicle should be between 2$m/s^2$ and -3.5$m/s^2$ \cite{openpilot_safety} (the exact thresholds set in PANDA with a more conservative design based on ISO 22179 \cite{ISO22179}), respectively, and only control commands that fit within the safe range can be sent to the simulation platform for execution.

\textbf{Human Driver Reactions:}
At Level 2 autonomy \cite{SAEroadmap}, drivers are required to monitor driving safety at all times and intervene in emergencies to prevent hazards. We implement a driver reaction simulator to evaluate the impact of driver interventions on longitudinal and lateral control.

As shown in Table \ref{tab:driverReaction}, when receiving an FCW alert, noticing unexpected acceleration, unsafe following distance (e.g., less than a vehicle length), or unsafe cruising speed (exceeding 10\% of the speed limit \cite{2to4safedistance}), or identifying a vehicle attempting to cut in from an adjacent lane, the driver initiates an emergency brake after the reaction time to avoid collisions. 
We implement the emergency braking following a driver brake behavior study \cite{gaspar2019driver}. For hazard mitigation in the lateral direction, if a lane departure warning (LDW) is triggered or the vehicle approaches a lane line too closely (predicted distance less than 0.5 meters), the driver steers the vehicle back to the center of the lane, following the reaction time. 

We use a fixed average driver reaction time of 2.5 seconds for these interventions unless otherwise specified in our simulations, based on various government-issued transportation policy guidelines \cite{dmvhandbook,nscguide} and existing literature\cite{sato2021dirty,zhou2024runtime}. To account for the variability in drivers’ reaction times and assess their impact on hazard mitigation, we also conduct additional experiments in Section \ref{sec:reactiontime}.

\begin{table}[t]
\caption{Driver Reaction Simulator.}
\label{tab:driverReaction}
\resizebox{\columnwidth}{!}{%
\begin{tabular}{@{}lll@{}}
\toprule
\textbf{Activation Condition} & \textbf{Driver Reaction} & \textbf{Reaction Time} \\ \midrule
FCW Alerts&
  \multirow{5}{*}{\begin{tabular}[c]{@{}l@{}}Emergency Brake \& \\ Zero Throttle \& \\ No changes in \\ the steering angle\end{tabular}} &
  \multirow{5}{*}{2.5 seconds} \\
Unsafe Cruise Speed     &                 &               \\ 
Unexpected Acceleration &                 &               \\
{Unsafe Following Distance} &&  \\
Other Vehicle Cutting in&& \\\midrule
Lane Departure Warning&
  \multirow{2}{*}{\begin{tabular}[c]{@{}l@{}}Steer vehicle back to\\ the center of the lane  \end{tabular}}&
  \multirow{2}{*}{2.5 seconds} \\
Unsafe Distance to Lane Lines&                 &               \\ \bottomrule
\end{tabular}%
}
\vspace{-2em}
\end{table}

\section{Experiments}
\begin{figure}[t]
    \centering
    \includegraphics[width=\linewidth]{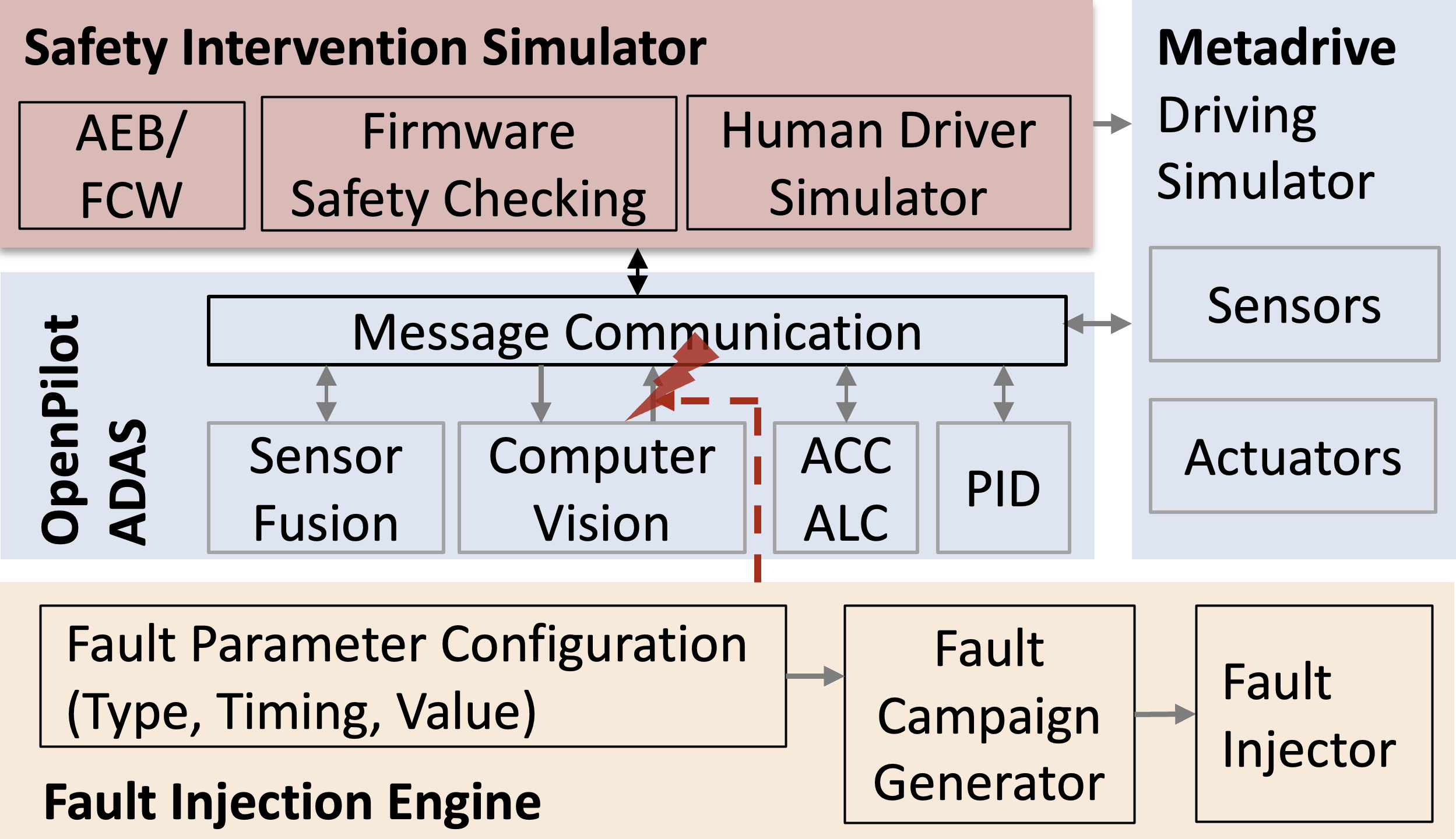}
    \caption{Overview of closed-loop simulation platform.}
    \label{fig:openpilotplatforml}
    \vspace{-1em}
\end{figure}

We develop a closed-loop simulation platform (see Fig. \ref{fig:openpilotplatforml}) to assess ADAS resilience and evaluate the effectiveness of various safety mechanisms. The platform integrates OpenPilot control software with the MetaDrive physical-world driving simulator and incorporates a fault injection engine alongside a safety intervention model as designed in Section \ref{sec:safety}. To address conflicts among safety interventions, we assign different priorities to the various safety mechanisms in our simulations, with AEB having the highest priority and safety checking the lowest \cite{maycointernational2024aeb} \cite{vtnews2024aeb}.

Experiments are conducted on two Ubuntu 20.04 LTS machines, one with an NVIDIA RTX 3070 graphics card and the other with an NVIDIA RTX 3090 graphics card. The platform utilizes OpenPilot v0.9.7 (the latest stable version) and MetaDrive 0.4.2.3. Each OpenPilot simulation comprises 10,000 time steps, with each step lasting approximately 10 ms, resulting in a total time of 100 seconds per simulation.


\subsection{Driving Scenarios}
\label{sec:scenario}
We simulate the following driving scenarios, which are identified as high-risk in the National Highway Traffic Safety Administration’s (NHTSA) pre-collision scenario typology report \cite{najm_pre-crash_2007}. In these scenarios, the ego vehicle cruises at 50 mph and approaches the lead vehicle from an initial distance of 60 or 230 meters. These distances are chosen to ensure the ego vehicle catches up with the lead vehicle on straight and curvy roads within our simulation.

\begin{itemize}
\item S1: The lead vehicle cruises at a constant speed (30 mph).
\item S2: The lead vehicle cruises at 30 mph and then accelerates to 40 mph.
\item S3: The lead vehicle cruises at 40 mph and then decelerates to 30 mph.
\item S4: The lead vehicle cruises at 30 mph and suddenly brakes to a stop due to an obstacle.
\item S5: The lead vehicle cruises at 30 mph, and another vehicle from the neighboring lane cuts into the ego vehicle's driving lane.
\item S6: Two lead vehicles cruise at a constant speed  (30 mph) in the same lane; then, the second lead vehicle (the one closer to the ego vehicle) changes lanes and moves into an adjacent lane.
\end{itemize}

\begin{figure}[t]
    \centering
    \includegraphics[width=\linewidth]{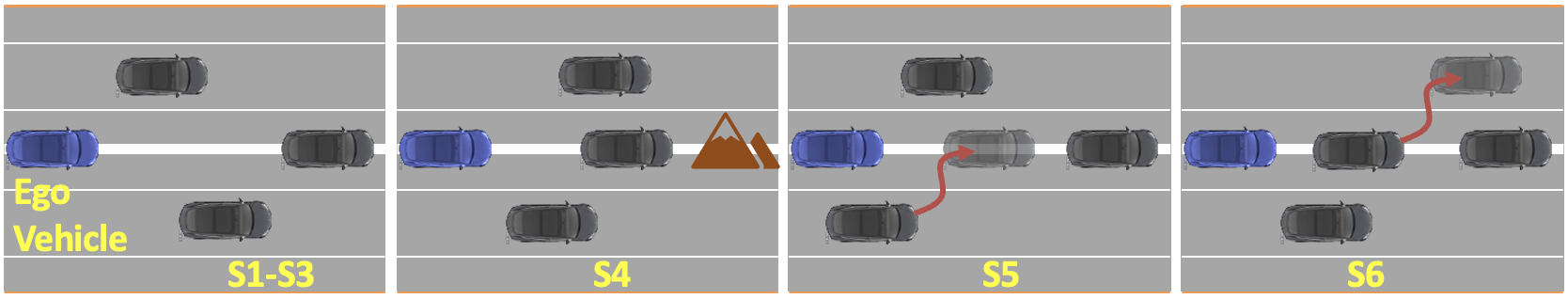}
    \caption{Driving Scenarios.}
    \label{fig:scenarios}
    \vspace{-1em}
\end{figure}

A visualization of each designed driving scenario is also shown in Fig. \ref{fig:scenarios}. In the MetaDrive simulator, we choose a dry highway map for all scenarios. The default environment is set to a bright morning with stable lighting and clear visibility. 

\subsection{Fault Injection}
\label{sec:faulttypes}

To simulate the effects of physical-world perception attacks on ACC and ALC (see Section \ref{sec:attackmodel}), we design a source-level fault injection (FI) engine capable of directly manipulating the outputs of the perception module by amounts consistent with those demonstrated in previous studies \cite{sato2021dirty,zhou2024runtime}. Specifically, the perception attack on ACC targets the DNN’s predictions of the lead vehicle’s position. To replicate this effect, the fault injection engine directly alters the predicted relative distance (RD) to the lead vehicle \cite{zhou2024runtime}. 
Similarly, the fault injection engine modifies the desired curvature predicted by the perception module to simulate sensor attacks on ALC \cite{sato2021dirty} (see Fig. \ref{fig:faulttype}).
The desired curvature, defined as the reciprocal of the turning radius, dictates the sharpness of the vehicle’s turns.
In this study, we consider fault injection targeting a single variable and multiple variables.

For each attack type listed in Table \ref{tab:FI}, the fault injection engine defines four parameters: (i) the target state variable, (ii) the magnitude of the error to inject, (iii) the trigger condition for the error, and (iv) the duration of the injected fault. Fault injections affecting the relative distance (RD) are triggered when the relative distance to the lead vehicle is less than 80 m, mimicking the perception of an adversarial patch on the rear of the lead vehicle by the ego vehicle. The injected fault values are set to 10 m, 15 m, and 38 m when the relative distance is within 80 m, 25 m, and 20 m, respectively, as reported in an existing work \cite{zhou2024runtime}. For attacks on ALC, the fault injection is activated when the ego vehicle crosses an adversarial patch on the road, introducing a 3\% deviation in curvature output predictions. This deviation results in a lateral path offset corresponding to a steering angle adjustment of up to 10 degrees, which falls within a reasonable range reported in the literature \cite{sato2021dirty}.
Each configuration is repeated 10 times, resulting in 360 simulations (3 fault types $\times$2 initial positions $\times$6 driving scenarios).

\begin{table}[t]
\centering
\caption{Fault Injection Parameters.}
\label{tab:FI}
\resizebox{\columnwidth}{!}
{%
\begin{tabular}{@{}llp{2.5cm}p{1.6cm}}
\toprule
\textbf{Type}                    & \textbf{Target Variable}    & \textbf{Attack Timing}   &\textbf{Attack Value}\\ 
                        \midrule
\multirow{3}{*}{Single} & Relative Distance   &  $RD$ < 80m          &38-10m\\ \cmidrule{2-4}
                        & \multirow{2}{*}{Desired Curvature} & {Ego vehicle} drives over road patch area           &\multirow{2}{*}{3\%} \\ \midrule
\multirow{3}{*}{Mixed}  & \multirow{3}{*}{RD \& Curvature}             &  $RD$ < 80m or ego vehicle drives across patch          &Same as above\\
\bottomrule 
\end{tabular}%
}
\vspace{-1em}
\end{table}

\subsection{Hazard and Accident}

We consider two types of hazards and accidents:
\begin{itemize}

 \item \textbf{A1}: Forward collision with the lead vehicle. \item \textbf{A2}: Driving out of the lane or colliding with side vehicles.  

 \item \textbf{H1}: Ego vehicle violates maintaining safety distance with the lead vehicle, which may result in A1.

 \item \textbf{H2:} Ego vehicle drives too close to the lane lines (e.g., 0.1m), which may lead to A2.

\end{itemize}



\subsection{Baselines}


Machine learning (ML) techniques have demonstrated significant advancements in hazard recovery for drones and unmanned aerial vehicles  \cite{dash2021pid, dash2024specguard}, yet their application to autonomous vehicles (AVs) remains relatively underexplored. To evaluate the effectiveness of ML-based automated methods in mitigating perception or sensor attacks against Level-2 ADAS, we develop a basic end-to-end ML baseline. This model takes as input the ego vehicle’s speed, relative distance to the leading vehicle, lane line positions, and historical gas and steering values from previous control cycles to directly predict the expected gas and steering outputs.

We train the model on fault-free data spanning 20 control cycles (0.2 seconds at a 100Hz control frequency) and explore various configurations of a two-layer LSTM model, including 256-128, 256-64, 256-32, 128-64, 128-32, and 64-32 hidden units. The best-performing LSTM model uses 128 and 64 hidden units. Since adding a third layer does not improve performance, we retain the two-layer model as the ML baseline.

\begin{algorithm}

\footnotesize
\DontPrintSemicolon
    \caption{Baseline ML-based Hazard Mitigation. }
    \label{alg:recovery}

\KwIn{Current state x(t) based on fault-free sensor data, threshold $\tau$, bias parameter $b(t)$} 
\KwOut{Control output to actuator $y(t)$}

\BlankLine

    $S(t) = 0$  \Comment{Initialize accumulated error} \\
    $b(t) = b_0 > 0$  \Comment{Initialize bias parameter} \\
    \While{Vehicle is operating}
    {
        $X_{t} \leftarrow [x(t-19),x(t-18),...,x(t)]$  \Comment{20 continuous frames} \\
        
        $Y_{t} \leftarrow [y(t-20),y(t-19),...,y(t-1)]$ \Comment{Historical outputs} \\
        $y_{OP} \leftarrow$ OpenPilot output \\
        $y_{ML} \leftarrow$ MLmodel.predict($X_t,Y_{t}$) \\
        $\delta \leftarrow | y_{ML} - y_{OP} | $  \\
        $S(t + 1) = max(0,S(t) + \delta - b(t))$ \Comment{Keep S(t) Non-negative}\\
        \uIf{$S(t + 1) > \tau$}
        {
            recovery\_mode $\leftarrow$ True
        }
        \uIf{recovery\_mode}
        {
            $y(t) \leftarrow y_{ML}$ \\
            \uIf{$\delta \leq b(t)$}
            {
                recovery\_mode $\leftarrow$ False \\
                $S(t) \leftarrow 0$  \Comment{Reset S(t) when exiting recovery mode}
            }
        }
        \uElse
        {
            $y(t) \leftarrow y_{OP}$
        }
    }

\end{algorithm}

As shown in Algorithm \ref{alg:recovery}, the mitigation mode is activated when the accumulated error between ML model predictions and OpenPilot controller outputs exceeds a preset threshold $\tau$~\cite{grigg2003use}. 
A bias parameter $b(t)$ > 0 is introduced to ensure that no error accumulates in $S(t)$ under normal conditions.

Under attack, we assume the ML model has access to fault-free input data from an independent or redundant sensor measurement, following the design of previous works \cite{dash2021pid} that isolate compromised sensors. The ML model then generates a mitigation output,  $y_{ML}$, which is executed by the actuator. The mitigation mode remains active until the error between ML outputs and OpenPilot outputs falls within threshold $b(t)$.



\subsection{Results}

\subsubsection{ADAS Driving Performance}
We first evaluate OpenPilot's ability to drive normally across six designed scenarios without any fault injection and present the results in Table \ref{tab:res-faultfree}. 

We observe that OpenPilot performs well in controlling the vehicle without accidents in most scenarios, including cut-in situations (S5). However, in Scenario S4, the OpenPilot fails to prevent a collision even in the absence of an attack, particularly when the lead vehicle brakes abruptly on a curve. The ego vehicle, still approaching at high speed without prior speed control before the curve, collides due to an insufficient emergency braking distance, despite triggering the FCW alarm.

\begin{table}[t]
\vspace{-1em}
\centering
\caption{Hardest Brake Value in Different Scenarios.}
\label{tab:res-faultfree}
\resizebox{\columnwidth}{!}{%

\begin{tabular}{@{}lllllll@{}}
\toprule 
\multirow{1}{*}{\textbf{Driving}}
  & \multirow{2}{*}{\textbf{Hazard}}
  & \multirow{2}{*}{\textbf{Accident}}
  & \multicolumn{1}{l}{\textbf{Following}}
  & \multicolumn{1}{l}{\textbf{Hard Brake}}
  & \multirow{1}{*}{min.}
  & \multirow{1}{*}{min.} \\
  \textbf{Scenario} & & & \textbf{Distance}(m) & \textbf{Value} & \textbf{TTC}(s) & {$\mathbf{t_{fcw}}$}(s) \\

 \midrule
 S1 &  1/20 & 0/20 &26.02 & 32.7\% & 5.70 & 4.42  \\
 S2 &  1/20 & 0/20 &29.15 & 15.7\% & 5.27 & 4.38  \\
 S3 &  2/20 & 1/20 &29.88 & 46.7\% & 3.71 & 4.39  \\
\textbf{S4} &  \textbf{10}/20 & \textbf{10}/20 &\textbf{23.72} & \textbf{86.7\%} & \textbf{0.85} & \textbf{3.24}  \\
 S5 &  2/20 & 1/20 &29.42 & 58.0\% & 2.33 & 3.90  \\
 S6 &  3/20 & 0/20 &28.15 & 30.3\% & 5.44 & 4.46  \\ 
 \bottomrule
\end{tabular}
}
\vspace{-2em}
\end{table}

\begin{figure}[t]
    \centering
    \begin{minipage}{0.31\columnwidth}
        \small
        \centering
        \includegraphics[width=\columnwidth]{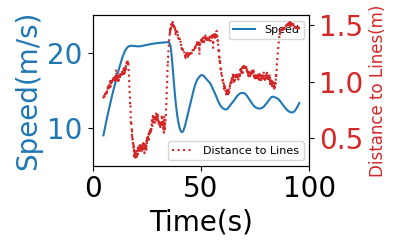}
        (S1)  
    \end{minipage}
    \hfil
    \begin{minipage}{0.31\columnwidth}
        \small
        \centering
        \includegraphics[width=\columnwidth]{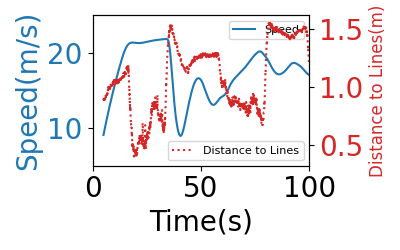}
        (S2)  
    \end{minipage}
    \hfil
    \begin{minipage}{0.31\columnwidth}
        \small
        \centering
        \includegraphics[width=\columnwidth]{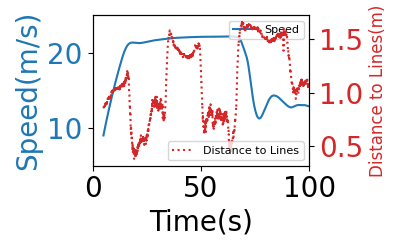}
        (S3)  
    \end{minipage}
    \hfil
    \begin{minipage}{0.31\columnwidth}
        \small
        \centering
        \includegraphics[width=\columnwidth]{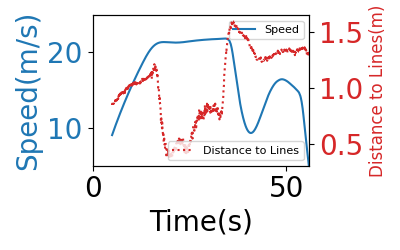}
        (S4)  
    \end{minipage}
    \hfil
    \begin{minipage}{0.31\columnwidth}
        \small
        \centering
        \includegraphics[width=\columnwidth]{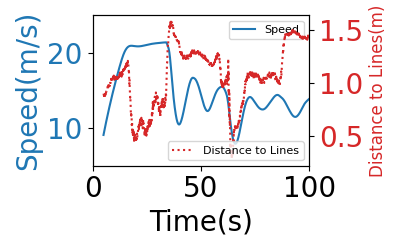}
        (S5)  
    \end{minipage}
    \hfil
    \begin{minipage}{0.31\columnwidth}
        \small
        \centering
        \includegraphics[width=\columnwidth]{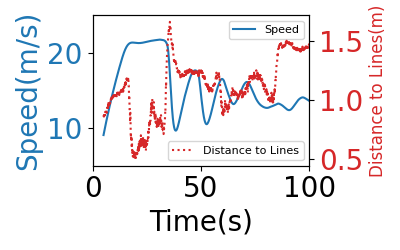}
        (S6)  
    \end{minipage}
    \caption{Speed and Distance to Lane Lines when Approaching LV.}
    \label{fig:speed-gas in 6 scenairos}
\end{figure}

Although OpenPilot maintains a larger following distance with the lead vehicle (LV) during a stable cruise stage compared to a previous version \cite{zhou2024runtime}, OpenPilot exhibits aggressive braking behavior when approaching the lead vehicle, even in the absence of any attacks. This overly aggressive braking, while not causing immediate hazards, can adversely affect ride smoothness and increase the risk of rear-end collisions in congested traffic conditions. As shown in Fig. \ref{fig:speed-gas in 6 scenairos}, taking Scenario 1 as an example, when the ego vehicle approaches the lead vehicle, the ego vehicle's speed suddenly drops from about 21.7m/s to 9.6m/s, a decrease of 55.8\% within 4.7 seconds, followed by similar speed fluctuations.

\begin{table}[t]
\caption{Minimal Distance to Lane Lines.}
\label{tab:alcdistance}
\resizebox{\columnwidth}{!}{%
\begin{tabular}{lllllll}
\toprule
\textbf{Driving Scenarios} & S1 & S2 & S3 & S4 & S5 & S6 \\ \midrule
\textbf{Distance to Lane Lines} (m) & 0.45 & 0.49 & 0.07 & 0.63 & 0.44 & 0.59\\ \bottomrule
\end{tabular}
}
\vspace{-2em}
\end{table}

Furthermore, ALC struggles to maintain the vehicle in the center of the lane, as evidenced by the minimal distance to lane lines, particularly during high-speed turns (see Table \ref{tab:alcdistance}). This increases the risk of side collisions and highlights insufficient coordination between the ALC and ACC modules. Additionally, the vehicle’s aggressive speed control on curves with excessive angles underscores the need for improved lateral and longitudinal stability to ensure safe driving performance.

\noindent\vsepfbox{
    \parbox{0.95\linewidth}{
       \textbf{Observation 1: OpenPilot fails to ensure safety in certain scenarios (e.g., S4) and exhibits aggressive braking when approaching a lead vehicle and inadequate lateral control for maintaining lane centering, highlighting the need for improved coordination between the ALC and ACC modules, particularly on curvy roads.}
    }
}

\subsubsection{ADAS Resilience under Attack}

\begin{table*}[t]
\caption{Fault Injection with or w/o Safety Interventions.}
\label{tab:res-FI}
\resizebox{\textwidth}{!}{%
\begin{tabular}{lllllllllllllll}
\toprule
            \textbf{Fault}  & \multicolumn{5}{c}{\textbf{Safety Interventions}} & \multicolumn{2}{c}{\textbf{Accidents}}& \textbf{Prevented} & \multicolumn{3}{c}{\textbf{Avg. Mitigation Time}(s)}  & \multicolumn{3}{c}{\textbf{Trigger Rate}}  \\ \cmidrule(lr){2-6} \cmidrule(lr){7-8}  \cmidrule(lr){10-12} \cmidrule(lr){13-15}
            
\textbf{Type}& {Driver}& Safety & AEB&AEB &ML&A1&A2&\textbf{Accident}&\multicolumn{1}{l}{AEB}&\multicolumn{1}{l}{Driver}&\multicolumn{1}{l}{Driver}&\multicolumn{1}{l}{AEB}&\multicolumn{1}{l}{Driver}&\multicolumn{1}{l}{Driver} \\
 & &Check & Comp.&Indep. &Model& & & & & Brake& Steering & & Brake & Steering \\ \midrule
 
\multirow{8}{*}{} & -& -& -&- &-& 82.50\% & 17.50\%& 0\%&\multicolumn{1}{l}{-}&\multicolumn{1}{l}{-}&\multicolumn{1}{l}{-}&\multicolumn{1}{l}{-}&\multicolumn{1}{l}{-}&\multicolumn{1}{l}{-} \\  

                  & $\checkmark$& $\checkmark$& -&- &-& 55.00\%& 0\%& 45.00\%&\multicolumn{1}{l}{-}&\multicolumn{1}{l}{3.19}&\multicolumn{1}{l}{0.97}&\multicolumn{1}{l}{-}&\multicolumn{1}{l}{51.67\%}&\multicolumn{1}{l}{6.70\%} \\  
                  & $\checkmark$& $\checkmark$& $\checkmark$&- &-& 49.17\%& 0\%& 50.83\%&\multicolumn{1}{l}{3.55}&\multicolumn{1}{l}{2.65}&\multicolumn{1}{l}{5.50}&\multicolumn{1}{l}{20.00\%}&\multicolumn{1}{l}{42.50\%}&\multicolumn{1}{l}{3.33\%} \\  
Relative                  & $\checkmark$& $\checkmark$& -&$\checkmark$ &-& 0\%& 0\%& \textbf{100\%}&\multicolumn{1}{l}{3.30}&\multicolumn{1}{l}{2.01}&\multicolumn{1}{l}{0.93}&\multicolumn{1}{l}{83.33\%}&\multicolumn{1}{l}{93.33\%}&\multicolumn{1}{l}{4.17\%} \\ 
 Distance                & -& -& $\checkmark$&- &-&  80.83\%&  0\%& 19.17\%&\multicolumn{1}{l}{3.46}&\multicolumn{1}{l}{-}&\multicolumn{1}{l}{-}&\multicolumn{1}{l}{19.17\%}&\multicolumn{1}{l}{-}&\multicolumn{1}{l}{-} \\ 
                  & -& -& -&$\checkmark$ &-& 0\%& 0\%& \textbf{100\%}&\multicolumn{1}{l}{3.26}&\multicolumn{1}{l}{-}&\multicolumn{1}{l}{-}&\multicolumn{1}{l}{100\%}&\multicolumn{1}{l}{-}&\multicolumn{1}{l}{-} \\ 
                  & $\checkmark$& -& -&- &-& 51.17\% & 0.83\% & 40.00\%&\multicolumn{1}{l}{-}&\multicolumn{1}{l}{2.72}&\multicolumn{1}{l}{3.00}&\multicolumn{1}{l}{-}&\multicolumn{1}{l}{66.00\%}&\multicolumn{1}{l}{3.33\%} \\ 
 & -& -& -& -& $\checkmark$& 1.67\%& 65.83\%& 32.5\%& -&- & -&- & -&-\\ \midrule
                  
\multirow{8}{*}{} & -& -& -&- &-& 0\%& 100\% & 0\%&\multicolumn{1}{l}{-}&\multicolumn{1}{l}{-}&\multicolumn{1}{l}{-}&\multicolumn{1}{l}{-}&\multicolumn{1}{l}{-}&\multicolumn{1}{l}{-} \\  
                  & $\checkmark$& $\checkmark$& -&- &-& 0\%& 54.17\%& 45.83\%&\multicolumn{1}{l}{-}&\multicolumn{1}{l}{4.26}&\multicolumn{1}{l}{0.92}&\multicolumn{1}{l}{-}&\multicolumn{1}{l}{30.00\%}&\multicolumn{1}{l}{40.83\%} \\  
                  & $\checkmark$& $\checkmark$& $\checkmark$&- &-& 0\%& 52.72\%& 47.27\%&\multicolumn{1}{l}{3.14}&\multicolumn{1}{l}{0.82}&\multicolumn{1}{l}{0.77}&\multicolumn{1}{l}{43.64\%}&\multicolumn{1}{l}{32.73\%}&\multicolumn{1}{l}{16.36\%} \\  
Desired   & $\checkmark$& $\checkmark$& -&$\checkmark$ &-& 0\%& 46.67\%& \textbf{53.33\%}&\multicolumn{1}{l}{3.52}&\multicolumn{1}{l}{3.15}&\multicolumn{1}{l}{0.13}&\multicolumn{1}{l}{42.50\%}&\multicolumn{1}{l}{39.17\%}&\multicolumn{1}{l}{14.17\%} \\ 
    Curvature     & -& -& $\checkmark$&- &-& 0\%& 60\% & 40.00\%&\multicolumn{1}{l}{3.55}&\multicolumn{1}{l}{-}&\multicolumn{1}{l}{-}&\multicolumn{1}{l}{40.83\%}&\multicolumn{1}{l}{-}&\multicolumn{1}{l}{-} \\ 
                  & -& -& -&$\checkmark$ &-& 0\%& 59.17\%& 40.83\%&\multicolumn{1}{l}{3.12}&\multicolumn{1}{l}{-}&\multicolumn{1}{l}{-}&\multicolumn{1}{l}{48.83\%}&\multicolumn{1}{l}{-}&\multicolumn{1}{l}{-} \\ 
                  & $\checkmark$& -& -&- &-& 0\%& 51.67\%& 48.33\%&\multicolumn{1}{l}{-}&\multicolumn{1}{l}{4.26}&\multicolumn{1}{l}{0.93}&\multicolumn{1}{l}{-}&\multicolumn{1}{l}{30.00\%}&\multicolumn{1}{l}{41.67\%} \\ 
 & -& -& -& -& $\checkmark$& 0\%& 60.00\%& 40.00\%& -& -& -& -& -&-\\ \midrule
                  
\multirow{8}{*}{} & -& -& -&- &-& 4.17\% & 95.83\% & 0\%&\multicolumn{1}{l}{-}&\multicolumn{1}{l}{-}&\multicolumn{1}{l}{-}&\multicolumn{1}{l}{-}&\multicolumn{1}{l}{-}&\multicolumn{1}{l}{-} \\  
                  & $\checkmark$& $\checkmark$& -&- &-& 7.50\% & 54.17\% & 38.33\% &\multicolumn{1}{l}{-}&\multicolumn{1}{l}{3.21}&\multicolumn{1}{l}{3.16}&\multicolumn{1}{l}{-}&\multicolumn{1}{l}{59.17\%}&\multicolumn{1}{l}{32.50\%} \\  
                  & $\checkmark$& $\checkmark$& $\checkmark$&- &-& 8.33\% & 41.67\% & 50.00\%&\multicolumn{1}{l}{3.35}&\multicolumn{1}{l}{2.93}&\multicolumn{1}{l}{2.85}&\multicolumn{1}{l}{12.50\%}&\multicolumn{1}{l}{80.83\%}&\multicolumn{1}{l}{72.50\%} \\  
                  & $\checkmark$& $\checkmark$& -&$\checkmark$ &-& 0\%& 48.33\%& 51.67\%&\multicolumn{1}{l}{3.68}&\multicolumn{1}{l}{3.20}&\multicolumn{1}{l}{0.87}&\multicolumn{1}{l}{41.67\%}&\multicolumn{1}{l}{36.67\%}&\multicolumn{1}{l}{12.50\%} \\ 
    Mixed           & -& -& $\checkmark$&- &-& 6.67\%& 67.50\%& 25.83\%&\multicolumn{1}{l}{0.05}&\multicolumn{1}{l}{-}&\multicolumn{1}{l}{-}&\multicolumn{1}{l}{25.83\%}&\multicolumn{1}{l}{-}&\multicolumn{1}{l}{-} \\ 
                  & -& -& -&$\checkmark$ &-& 0\%& 58.33\%& 41.67\%&\multicolumn{1}{l}{3.62}&\multicolumn{1}{l}{-}&\multicolumn{1}{l}{-}&\multicolumn{1}{l}{43.33\%}&\multicolumn{1}{l}{-}&\multicolumn{1}{l}{-}\\ 
                  & $\checkmark$& -& -&- &-& 8.33\%& 22.50\%& \textbf{69.17\%}&\multicolumn{1}{l}{-}&\multicolumn{1}{l}{3.18}&\multicolumn{1}{l}{0.93}&\multicolumn{1}{l}{-}&\multicolumn{1}{l}{64.17\%}&\multicolumn{1}{l}{32.50\%} \\ 
 & -& -& -& -& $\checkmark$& 0\%& 76.92\%& 23.08\%& -& -& -& -& -&-\\ \bottomrule

\end{tabular}
}
 \vspace{-2em}
\end{table*}

We assess the resilience of ADAS against attacks (see Table \ref{tab:FI}) without any safety intervention. We present the results in Table \ref{tab:res-FI}. We see in Table \ref{tab:res-FI} that accidents happen in all the simulations, indicating the insufficiency of OpenPilot in tolerating adversarial perception attacks and the need for safety mechanisms to improve safety. Additionally, 17.5\% of out-of-the-lane (A2) accidents are caused by fault injection against ACC, resulting from overspeeding on curvy roads and lane changes to avoid forward collisions in cut-in scenarios. 
For mixed attacks, more A2 accidents occur than A1 accidents due to the shorter time needed to trigger accidents in the latter direction, highlighting the severe vulnerability of ALC to perception attacks.

\begin{figure}[t]
    \centering
    \includegraphics[width=0.45\linewidth]{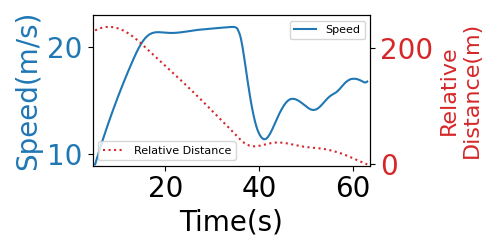}
    \caption{Speed and Relative Distance under Fault Injection.}
    \label{fig:speed-RD}
    \vspace{-2em}
\end{figure}

We also observe that OpenPilot fails to recognize the lead vehicle when the relative distance (RD) is short. As shown in Fig. \ref{fig:speed-RD}, the ego vehicle continues to approach the lead vehicle using the tampered input. However, once the ego vehicle gets within a certain range, such as 2 meters, OpenPilot is unable to detect the lead vehicle through the camera. This failure causes the ego vehicle to accelerate, resulting in a collision.

\noindent\vsepfbox{
    \parbox{0.95\linewidth}{
       \textbf{Observation 2: OpenPilot is unable to tolerate perception attacks against ACC and ALC and fails to recognize the front vehicle at a short distance.}
    }
}

\subsubsection{Evaluation of Safety Interventions}
Table \ref{tab:res-FI} also shows the results of fault injection experiments with various configurations of safety interventions.
We see in Table \ref{tab:res-FI} that the designed safety mechanisms can prevent 19.17-100\% accidents across different safety configurations and fault types, demonstrating safety interventions’ key role in mitigating ADAS perception attacks.

Among the tested mechanisms, AEB utilizing an independent data source consistently outperforms other strategies in preventing forward collisions, achieving an accident prevention rate of up to 100\%. This underscores the effectiveness of AEB in addressing immediate forward hazards. However, when AEB uses compromised data, the accident prevention rate drops significantly to 19.17\%, highlighting the importance of designing an AEB with inputs from an independent sensor or a fault-resilient data stream.
While AEB is not designed to address hazards or accidents in the lateral direction, we observe that it prevents up to 40.83\% of A2 accidents caused by attacks on ALC. This occurs because the ego vehicle's aggressive acceleration toward the lead vehicle activates AEB, stopping the ego vehicle from driving out of the lane.

We also observe human driver intervention, including braking and steering,  prevents a significant portion of accidents across different fault types (40-69.17\%), indicating the important role of driver in ensuring safety in addition to advanced safety features (e.g., AEB, FCW) when driving $\text{Level-2}$ AVs~\cite{SAEroadmap}.
However, the average driver reaction time (2.5 seconds) introduces a critical delay, which impacts the effectiveness of interventions in fast-evolving situations. For example, in relative distance {attacks},
where rapid deceleration is required to prevent collisions, the delayed response limits the accident prevention rate compared to AEB (40\% vs. 100\%). Conversely, for curvature attacks leading to lane deviations, driver steering interventions demonstrate higher efficacy. 

\noindent\vsepfbox{
    \parbox{0.95\linewidth}{\textbf{Observation 3: AEB and driver intervention can prevent accidents in both driving and lateral directions and are critical to ensuring driving safety.}
}}


For the mixed fault type, human intervention achieves the highest hazard prevention rate at 69.17\%. However, when combined with AEB, the prevention rate drops to 50–51.67\%. This decline is primarily due to the prevalence of lateral-direction accidents (A2), where AEB underperforms compared to human intervention. Since AEB has the highest priority in the safety hierarchy, it overrides human inputs, leading to more unsuccessful mitigation cases. These findings underscore the challenges of prioritizing safety mechanisms to prevent conflicts across diverse driving and fault scenarios. While AEB is critical for immediate risk mitigation, human intervention remains vital for addressing complex or prolonged faults. Driver responses provide a complementary layer of safety, particularly when automated systems are disabled or compromised.

\noindent\vsepfbox{
    \parbox{0.95\linewidth}{\textbf{Observation 4: Better coordination of safety mechanisms at different levels is needed to resolve conflicts and ensure safety under complex attacks.}
}}

\subsubsection{Evaluation with Different Driver Reaction Times}
\label{sec:reactiontime}
Although the designed safety mechanisms mitigate up to 53.33\% of the accidents caused by attacks against ALC, they still fail to prevent nearly half of the accidents. This is due to the short mitigation time in the lateral direction and the critical delay in driver reaction. 
To account for the variability in driver’s reaction time and further assess their impact on hazard mitigation outcomes, we rerun the simulations with reaction times varying between 1.0s and 3.5s \cite{BrakeReactionTime} \cite{MAKISHITA2008567}, using the same scenario and fault settings introduced in Sections \ref{sec:scenario} and \ref{sec:faulttypes}, while enabling only driver interventions. This results in 360 simulations for each reaction time configuration. 
Table \ref{tab:driverrectime} demonstrates that the success rate of driver intervention increases when reaction time is shorter than 2 seconds, emphasizing the importance of staying alert while driving.

\begin{table}[t]
\caption{Prevention Rate vs. Driver Reaction Time. }
\label{tab:driverrectime}
\resizebox{\columnwidth}{!}{%
\begin{tabular}{lllllll}
\toprule
\multirow{2}{*}{\textbf{Fault Type}} & \multicolumn{6}{c}{\textbf{Driver Reaction Time(s)}}                                                                            \\ \cmidrule{2-7} 
                  & \multicolumn{1}{l}{1.0} & \multicolumn{1}{l}{1.5} & \multicolumn{1}{l}{2.0} & \multicolumn{1}{l}{2.5} &  3.0 &3.5\\ 
                  \midrule
               
                  Relative Distance& \multicolumn{1}{l}{53.33\%} & \multicolumn{1}{l}{55\%} & \multicolumn{1}{l}{55\%} & \multicolumn{1}{l}{40\%} &  43.33\% & 41.67\% \\ 
                  Desired Curvature& \multicolumn{1}{l}{77.50\%} & \multicolumn{1}{l}{55.83\%} & \multicolumn{1}{l}{58.11\%} & \multicolumn{1}{l}{48.33\%} &  52.50\%& 40.00\%\\ 
                  Mixed& \multicolumn{1}{l}{70.83\%} & \multicolumn{1}{l}{70.00\%} & \multicolumn{1}{l}{68.33\%} & \multicolumn{1}{l}{69.17\%} & 60.83\%& 53.33\% \\ \bottomrule
\end{tabular}
}
\vspace{-2em}
\end{table}

\noindent\vsepfbox{
    \parbox{0.95\linewidth}{
    \textbf{Observation 5: Attacks against ALC cannot be easily mitigated while an alert driver may improve the success rate of accident prevention.}
}}

\subsubsection{Evaluation with Various Environmental Conditions} Environmental conditions impact hazard mitigation in two ways: lighting affects visibility and perception, while weather conditions, such as rain, impact road friction. In MetaDrive, there is no direct interface to modify environmental conditions, and adjusting lighting using third-party renderers degrades OpenPilot’s functionality under normal operation. Therefore, we can only simulate varying road friction by modifying friction parameters to represent wet or icy conditions. To assess the effect of different weather conditions on safety mechanisms for hazard mitigation, we rerun all simulations across all fault types and scenarios.

Table \ref{tab:weather} shows that the hazard mitigation rate declines as road friction decreases. However, the safety mechanisms maintain a similar prevention rate at 50\% road friction (e.g., heavy rain \cite{SupercarsTraction}), indicating a certain level of robustness in mitigating hazards under varying weather conditions. When road friction is reduced to 25\% 
(e.g., icy road \cite{rasol2023weather}), we observe a significant drop in the hazard mitigation rate against attacks targeting desired curvature, highlighting the increased vulnerability of ALC in severe weather conditions.






\begin{table}[h]
\vspace{-0.5em}
\centering
\caption{Hazard Prevention Rate vs. Road Friction.}
\label{tab:weather}
\begin{tabular}{lllll}
\toprule
      \multirow{2}{*}{\textbf{Fault Type}} & \multicolumn{4}{c}{\textbf{Road Friction}}  \\ \cmidrule{2-5}
      
      &Default & 25\% off & 50\% off & 75\% off \\ \midrule
Relative Distance& 50.83\% & 51.65\% & 47.50\% & 43.33\% \\
Curvature/Lateral        & 47.27\% & 44.17\% & 45.83\% & 18.33\% \\
\bottomrule
\end{tabular}

\footnotemark{Enabled Safety Intervention: Driver, Safety Check, AEB Compromised}
\vspace{-\baselineskip}

\end{table}

\subsubsection{Comparison to ML-based Mitigation}
From Table \ref{tab:res-FI} we see that the ML-based baseline achieves a hazard prevention rate of 23.08\%–40.00\% across different fault types, which, while reasonable, remains lower than driver intervention and AEB. 
While the ML model prevented nearly all A1 accidents caused by relative distance attacks, it introduced new A2 accidents, highlighting its inadequate performance in vehicle centering and the need for further efforts in designing and training a more advanced ML model.

\noindent\vsepfbox{
    \parbox{0.95\linewidth}{\textbf{Observation 6:  Basic safety mechanisms (e.g., AEB) or human intervention may be more effective than certain automated ML-based mitigation methods in preventing accidents in ADAS in our simulated scenarios.}
}}

\section{Threats to Validity}
This study is conducted solely in simulation, leaving the real-world impact of safety mechanisms on perception attacks against ADAS in actual vehicles uncertain. 
Further, the driver reaction model relies on fixed rules and thresholds, which may not fully represent human behavior or distribution of reaction time during emergencies. 
We try to mitigate this weakness by developing a realistic testbed 
and examining various driver reaction times.
Testing on actual vehicles and modeling more complex driver reactions are directions of future work.


\section{Related Work}

Existing work has explored the vulnerability of autonomous driving and AVs against faults or attacks \cite{10.1145/3372297.3423359} targeting Lidar~\cite{lidar_2023}, GPS \cite{shen2020drift}, radar \cite{komissarov2021spoofing}, camera  \cite{sato2021dirty,nassi2019mobilbyeattackingadascamera}, perception model \cite{schmedding2024strategic,SchmeddingYJS24,YangNJS21}, CAN bus  \cite{zhou2022strategic}, multi-sensor fusion~\cite{10.1145/3636534.3649372}, object tracking \cite{jha2020ml}, or controller \cite{zhou2023hybrid}. 
However, most works do not account for the existing safety mechanisms.
{Few studies on safety interventions have focused solely on a single ADAS feature, such as ACC \cite{zhou2024runtime}, without accounting for the interactions between multiple safety mechanisms.}
Additionally, some studies have evaluated the resilience and robustness of autonomous systems \cite{tuncali2018simulation,tian2018deeptest,ben2016testing,abdessalem2018testing,gambi2019automatically,RahmanADASRA,10.1145/3508352.3561111,10583859}. However, these efforts often focus on individual components, such as AEB \cite{RahmanADASRA} or FCW~\cite{ma2021sequential}, and lack a comprehensive, system-wide testing approach.
A comparison of our work with other existing work is shown in Table~\ref{table:related}.

\begin{table}[h!]
\vspace{-0.5em}
\caption{Comparison with Existing Work.}
\label{table:related}

\resizebox{\columnwidth}{!}{%
\begin{tabular}{llllll}

\toprule
\multirow{2}{*}{\textbf{Work}} & \multirow{2}{*}{\textbf{Attack Vector}} & \multirow{2}{*}{\textbf{Target}}  & \multirow{2}{*}{\textbf{AEB}}  & \textbf{Driver} & \textbf{Autonomy}  \\ 
&   && & \textbf{Intervention} & \textbf{Level}\\
 \midrule
 ~\cite{zhou2022strategic} &  \multirow{2}{*}{Control Commands}  & ACC, ALC & N & Y & L2  \\
 ~\cite{10478772} & & ACC & Y & N & L2  \\ \cline{1-6}
 ~\cite{RahmanADASRA} & \multirow{5}{*}{Perception Input} & LiDAR, Camera & Y & N & N/A  \\
 ~\cite{10583859} &  & ALC & N & N & L2  \\
 
 ~\cite{10.1145/3636534.3649372} &  & LiDAR, Camera, Radar & N & N & N/A  \\
 ~\cite{10.1145/3372297.3423359} &  & ADAS & N & N & L2  \\
 Ours &   & ACC, ALC & \textbf{Y} & \textbf{Y} & L2  \\
 \bottomrule
\end{tabular}
}
\vspace{-1.5em}
\end{table}

\section{Conclusion}
This paper systematically evaluates the resilience of an open-source ADAS to adversarial patch attacks and the effectiveness of safety interventions in mitigating hazards and ensuring driving safety. Our findings show that OpenPilot is highly susceptible to these attacks targeting ACC and ALC, often failing to detect the front vehicle at close range and displaying unsafe, aggressive speed control even in benign conditions. 
Lateral attacks remain challenging to mitigate, though highly alert drivers achieve better prevention rates. Further, the results emphasize the potential of AEB to prevent lateral accidents, the importance of independent sensors for AEBS, and the urgent need to address conflicts among safety features to enhance overall system reliability and resilience.



\section*{Acknowledgment}
This work was partially supported by the National Science Foundation (NSF) under Grants 2402940, 2402941, 2402942, 2410856, and CCI HC-3Q24-047.

\bibliographystyle{IEEEtran}
        
\bibliography{reference}

\end{document}